\documentclass[
aps,
prl,
amsmath,
amssymb,
aps,
superscriptaddress,
twocolumn,
]{revtex4-2}
\usepackage[T1]{fontenc} 
\usepackage{graphicx}
\usepackage{dcolumn}
\usepackage{bm}
\usepackage{hyperref}
\usepackage{appendix}
\usepackage{xcolor}

\usepackage[normalem]{ulem} 

\begin{document} 

\title{Anomalous fluctuations of Bose-Einstein condensates in optical lattices}

\author{Zahra Jalali-Mola}
\affiliation{ICFO - Institut de Ciencies Fotoniques, The Barcelona Institute of Science and Technology, 08860 Castelldefels, Barcelona, Spain}

\author{Niklas Käming}
\affiliation{Institute for Quantum Physics, University of Hamburg, 22761 Hamburg, Germany}
\affiliation{The Hamburg Centre for Ultrafast Imaging, 22761 Hamburg, Germany}

\author{Luca Asteria}
\affiliation{Department of Physics, Graduate School of Science, Kyoto University, Kyoto 606-8502, Japan}

\author{Utso Bhattacharya}
\affiliation{ICFO - Institut de Ciencies Fotoniques, The Barcelona Institute of Science and Technology, 08860 Castelldefels, Barcelona, Spain}
\affiliation{Institute for Theoretical Physics, ETH Zurich, Zurich, Switzerland}

\author{Ravindra W. Chhajlany}
\affiliation{Institute of Spintronics and Quantum Information, Faculty of Physics,Adam Mickiewicz University,
61-614 Poznań, Poland}

\author{Klaus Sengstock}
\affiliation{Institute for Quantum Physics, University of Hamburg, 22761 Hamburg, Germany}
\affiliation{The Hamburg Centre for Ultrafast Imaging, 22761 Hamburg, Germany}

\author{Maciej Lewenstein}
\affiliation{ICFO - Institut de Ciencies Fotoniques, The Barcelona Institute of Science and Technology, 08860 Castelldefels, Barcelona, Spain}
\affiliation{ICREA, Pg. Llu\'is Companys 23, 08010 Barcelona, Spain}

\author{Tobias Grass}
\email {tobias.grass@dipc.org}
\affiliation{DIPC - Donostia International Physics Center, Paseo Manuel de Lardiz{\'a}bal 4, 20018 San Sebasti{\'a}n, Spain}
\affiliation{Ikerbasque - Basque Foundation for Science, Maria Diaz de Haro 3, 48013 Bilbao, Spain}

\author{Christof Weitenberg}
\email {christof.weitenberg@tu-dortmund.de}
\affiliation{Department of Physics, TU Dortmund University, 44227 Dortmund, Germany}

\begin{abstract}
Fluctuations are fundamental in physics and important for understanding and characterizing phase transitions. In this spirit, the phase transition to the Bose-Einstein condensate (BEC) is of specific importance. Whereas fluctuations of the condensate particle number in atomic BECs have been studied in continuous systems, experimental and theoretical studies for lattice systems were so far missing. Here, we explore the condensate particle number fluctuations in an optical lattice BEC across the phase transition in a combined experimental and theoretical study. We present both experimental data using ultracold $^{87}$Rb atoms and numerical simulations based on a hybrid approach combining the Bogoliubov quasiparticle framework with a master equation analysis for modeling the system. We find strongly anomalous fluctuations, where the variance of the condensate number $\delta N_{\rm BEC}^2$ scales with the total atom number as $N^{1+\gamma}$ with an exponent around $\gamma_{\rm theo}=0.74$ and $\gamma_{\rm exp}=0.62$, which we attribute to the 2D/3D crossover geometry and the interactions. Our study highlights the importance of the trap geometry on the character of fluctuations and on fundamental quantum mechanical properties.
\end{abstract}

\date{\today}

\maketitle

\begin{figure}
    \centering
    \includegraphics[width=1\linewidth]{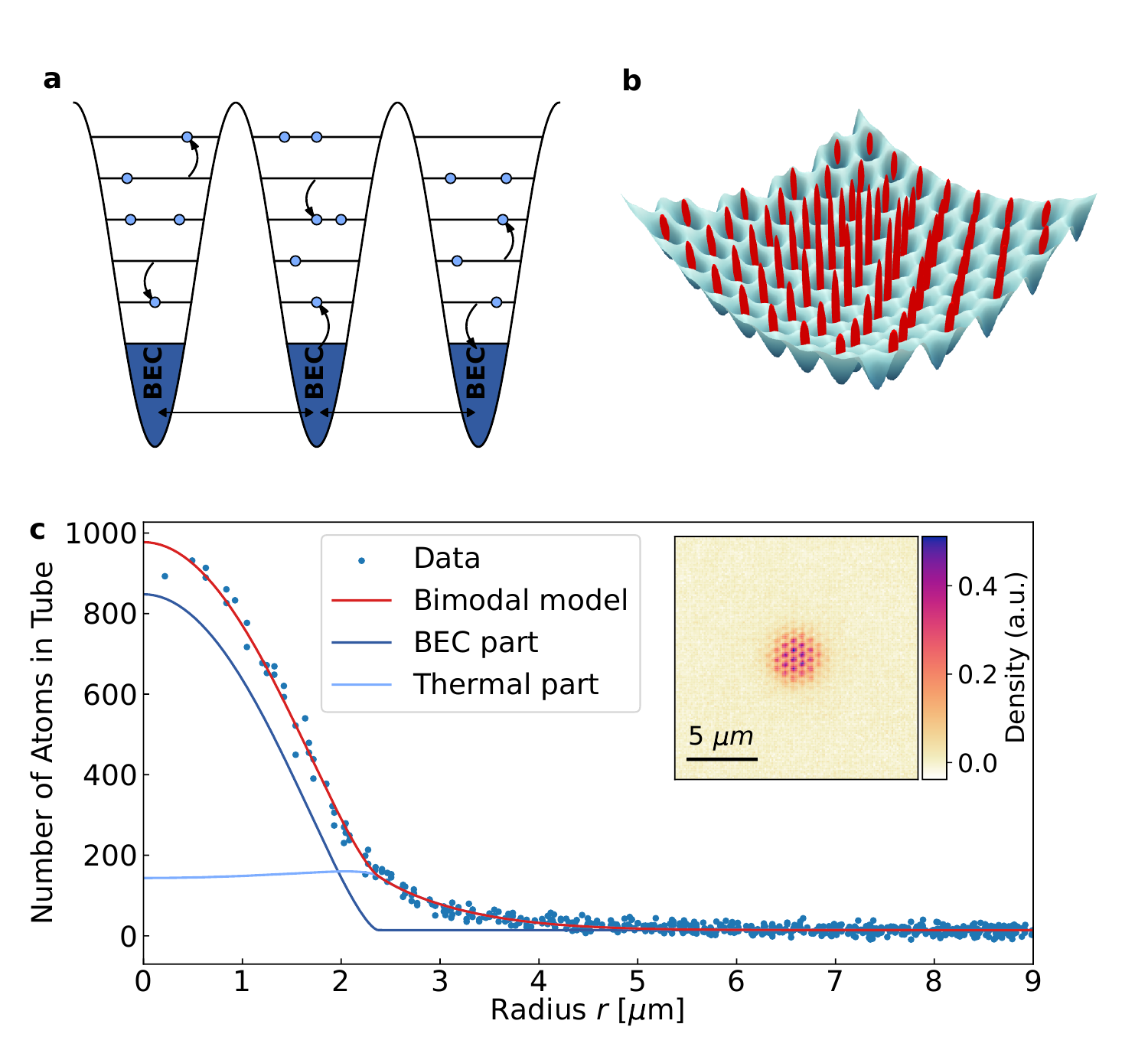}
    \caption{Trap geometry and density profiles of the BECs. \textbf{(a)} The fluctuations arise due  to continuous particle exchange between different energy levels.  \textbf{(b)} The trap geometry consists of a triangular lattice of tubes within a harmonic confinement.   
    \textbf{(c)} Radial density profile of tube occupations of a typical BEC along with the fitted bimodal distribution, consisting of the density of the BEC $n_\textnormal{BEC}$ and the density of the thermal cloud $n_\textnormal{th}$. The inset shows the density observed with the matter-wave microscope. This image was obtained for an evaporation frequency $f_\textnormal{evap} = 95.0$ kHz and yields fit parameters of BEC particle number $N_\textnormal{BEC} = 13300 \pm 150$, fugacity $z = 2.13 \pm 0.14$, temperature $T = 115 \pm 3$ nK and overall particle number $N_\textnormal{tot}=24100 \pm 600$.}
    \label{fig:system}
\end{figure}

\textit{Introduction.}
Fluctuations are fundamental to the structure of physical reality. They provide information about the behavior of systems at different scales, from the quantum realm\cite{Campisi2011} to the cosmos~\cite{Harrison1967}, and are essential for both the theoretical understanding and practical applications of physics. The fluctuation-dissipation theorem establishes a deep connection between the fluctuations in a system at equilibrium and its response to external perturbations~\cite{Kubo1966}. Fluctuations near critical points can drive phase transitions, creating ordered phases of matter through a mechanism known as order-by-disorder~\cite{Villain1980,Bergman2007,Turner2007,Barnett2012}. A testbed for studying fluctuations in a quantum system are atomic Bose-Einstein condensates in which a macroscopic number of bosonic particles accumulate in the ground state. This phase of matter, predicted in 1925, was for atomic gases first experimentally realized 70 years later, using magnetic traps to confine and cool alkali atoms to extremely low temperatures~\cite{BEC_exp1,BEC_epx2}. While conceptually simple, the condensate features interesting behavior, such as long-range phase coherence or superfluid flow~\cite{Book_pitaevskii}. The number fluctuations of Bose-Einstein condensates exhibit particularly subtle properties that have spurred theoretical discussions and investigations for many decades, see the review articles in Refs.~\cite{Kocharovsky2006,Kruk2025}. More recently, fluctuations in atomic Bose-Einstein condensates have also become subject of experimental investigations~\cite{Kristensen2019,Christensen2021,Vibel2024}. 

Notably, even in the theoretically most tractable case of an ideal Bose gas, the number fluctuations of the condensate behave in a very intriguing way: different thermodynamic ensembles strongly disagree regarding the amount of fluctuations in a non-interacting gas \cite{Ziff1977, Navez1997, Wilkens1997,Holthaus1998,Kocharovsky2006,Kruk2023}, while being consistent regarding the mean value of the condensate fraction. The grand-canonical ensemble exhibits a particularly dramatic behavior, which is often referred to as the grand-canonical catastrophe: condensate fluctuations $\delta N_{\rm BEC}$ are of the order of the particle number $N$\cite{Ziff1977}. In contrast, the  microcanonical and canonical ensembles yield the same amount of fluctuations in 1D \cite{Wilkens1997,Holthaus1998}, whereas in a 3D trapped gas, the two ensembles agree only with respect to the scaling behavior, and the absolute amount of fluctuations is suppressed in the microcanonical ensemble~\cite{Navez1997,Weiss1997}. The suppression of fluctuations in the microcanonical ensemble has also been seen experimentally, with the additional feature of interaction-induced modifications~\cite{Christensen2021}. The condensation of photons~\cite{Weiss2016,Crisanti2025} has opened the door towards experimental investigation of the scenario given by an ideal gas in the grand-canonical ensemble, realized by an effective particle exchange between the photon gas and a dye reservoir by absorption and emission processes, although the finite size of the dye reservoir needs to be taken into account. The grand canonical BEC fluctuations have been measured in this system via second order coherence of the photons \cite{Schmitt2014}.

Apart from the grand-canonical catastrophe, the scaling of the condensate number fluctuations with the system size is found to deviate from Gaussian behavior $\delta N_{\rm BEC}^2 \propto N$ also in other ensembles. In this context, a tremendously important role is played by the geometry of the system. In a box-shaped potential, the single-particle energies $\epsilon_i \propto k_i^\sigma$ disperse quadratically, $\sigma=2$, whereas in a harmonic trap potential the dispersion is linear, $\epsilon_i \propto n_i^\sigma$ with $\sigma=1$, where  $k_i$ and $n_i$ denote the quantum numbers in the respective geometry. Together with the dimensionality $d$ of the system the dispersive behavior $\sigma$ has crucial consequences on the phenomenon of Bose-Einstein condensation~\cite{deGroot1950,Weiss1997,Kruk2025}, and in particular on the scaling of condensate fluctuations: While for $d<\sigma$ condensation does not even occur at all, the condensate of ideal bosons exhibits normal fluctuations for $d>2\sigma>0$, with a scaling behavior $\delta N_{\rm BEC}^2 \propto N \tilde T^ {d/\sigma}$ at low temperature, where $\tilde T=T/T_{\rm c}$ denotes reduced temperature, that is, the temperature $T$ in units of the critical temperature $T_{\rm c}$. Such scenario corresponds, for instance, to a 3D Bose gas in a harmonic trap. In contrast to this, for $\sigma<d<2\sigma$, the scaling of fluctuation is anomalous, $\delta N_{\rm BEC}^2 \propto N^{2\sigma/d} \tilde T^2$. Accordingly, for the 3D box potential, the scaling is $\delta N_{\rm BEC}^2 \propto N^{1+\gamma}$ with an an exponent $\gamma=1/3$ \cite{Giorgini1998}.

For the interacting Bose gas, the scaling behavior of fluctuations has given rise to controversial results, as exact results for interacting systems are not available, and the predictions sensibly depend on the chosen approximations~\cite{Xiong2002} and finite-size effects. For instance, the Bogoliubov treatment ~\cite{Giorgini1998} as well as number-conserving approaches~\cite{Kocharovsky2000,Idziaszek2005} predict anomalous scaling for a harmonically trapped 3D gas, whereas a perturbative approach found normal scaling~\cite{Idziaszek1999}. Ref.~\cite{Yukalov2005} has argued that anomalous scaling in the Bogoliubov approach is due to inconsistent higher-order terms.
For a recent review of this controversy, see also Ref.~\cite{Kruk2025}.
Experiments with interacting ultracold atoms in an elongated 3D harmonic trap found a scaling exponent of $\gamma=0.134(5)$  \cite{Christensen2021}, which is reduced compared to the prediction $\gamma=1/3$ for the interacting gas in the 3D harmonic trap \cite{Giorgini1998,Meier1999} due to the dimensional effect of the elongated trap.

In contrast to the vast amount of literature on condensate fluctuations in continuous systems, very little attention has been paid to lattice systems~\cite{Dutkiewicz2025}, despite the general great importance of optical lattices for ultracold atoms~\cite{Maciej_book}. The present work closes this gap of knowledge, and investigates, both experimentally and theoretically, the case of bosonic atoms in an optical lattice potential. In addition, a three-dimensional (3D) harmonic trap potential confines the atoms, with the trap along the perpendicular direction being much weaker than within the lattice plane. This combination of traps gives rise to a triangular array of tubes, as illustrated in Fig.~1b. A BEC in an array of tubes was shown to exhibit the physics of 2D systems concerning vortex proliferation and short-range coherence \cite{Schweikhard2007,Bruggenjurgen2024}
but the behavior of condensate fluctuations in such  geometry, also known as 2D/3D crossover, has not been studied.
Here we show that such a system exhibits
strongly anomalous scaling of the condensate fraction fluctuations with the particle number. 
On the experimental side, we use a matter-wave microscope for precision thermometry from high-resolution density measurements of ultracold $^{87}$Rb atoms \cite{Asteria2021} allowing to extract the condensate fraction for each image. On the theory side, we use a hybrid approach combining the Bogoliubov quasiparticle framework with a master equation analysis~\cite{Scully_PRL1999,Scully_PRA2000,Scully_2006,scully_2010,scully_PRL}. 
We find $\gamma_{\rm theo}= 0.74$ for the numerics and $\gamma_{\rm exp}= 0.62$ for the experimental data. These values match well, and they are in between the expected values for non-interacting systems with the density of states of a truly 2D lattice ($\gamma=1$) and a truly 3D lattice ($\gamma=1/3$) \cite{Kocharovsky2006}. The fluctuations follow a curve as a function of the reduced temperature, which has a pronounced peak below $T_{\rm c}$, which further shifts down with increasing interactions.

\textit{Experimental setup.}
Our experiments start with BECs of $^{87}\rm{Rb}$ atoms prepared in a magnetic trap in the $F=2,\:m_F=2$ state. The trap is radially symmetric in the x,~y plane with a high trapping frequency of $\omega_{xy}=2\pi\times305$~Hz and a trapping frequency of $\omega_z = 2\pi \times 29$~Hz in the z-direction. 
We load into a triangular optical lattice formed by lattice beams with a wavelength of $\lambda = 1064$~nm, yielding a lattice constant $a_{\rm lat} = 2\lambda/3 = 709$~nm, and the recoil energy $E_{\rm rec} = \frac{h^2}{2m \lambda^2}$ with Planck's constant $h$ and the atomic mass $m$. 
We work with a lattice depth of $1~E_{\rm rec}$ 
\cite{Asteria2021,Kosch2022}, which yields a tunneling energy of $J = h\times 12$~Hz. 

To map out the phase transition, we prepare the system at different temperatures and atom numbers by varying the end point of the radio-frequency evaporation ramp in the magnetic trap before the loading into the lattice between 95 and 119~kHz, with the resonance frequency to the $m_F=1$ state being $\lesssim 90$~kHz. For high-resolution imaging of the real-space profiles with single-tube resolution, we employ the matter-wave microscope technique described in \cite{Asteria2021,Brandstetter2024}. We first ramp the lattice to $6~E_\textnormal{rec}$ ($J/h\sim10^{-3}$~Hz) to freeze the density distribution and wait for 12~ms to let the system dephase, in order to avoid interaction effects during the matter-wave protocol due to density peaks. We then ramp the in-plane trap frequency to $\omega_\textnormal{pulse}=2\pi\times390$~Hz for a large matter-wave magnification of 44. The protocol consists of a quarter-period pulse in the harmonic trap initialized by switching off the optical lattice and a subsequent free evolution by 18 ms. Finally, we perform absorption imaging with an optical magnification of 2 on a CCD camera.

We determine the atom number, temperature and condensate fraction from each experimental image by fitting a semi-ideal bimodal model \cite{SupMat} to the density profiles, integrated over the Wigner-Seitz cells (Fig.~1c inset). The bimodal fit contains the BECs fraction and the thermal fraction as well as the repulsive interaction of the BEC onto the thermal gas. We note that the high-resolution imaging of the density distribution via the matter-wave microscope allows precision thermometry of single images and precise modeling of the condensate fraction. This enables us to study the BEC fluctuations without the stabilization of the atom number using non-destructive imaging near the end of the evaporation ramp as in Ref.~\cite{Kristensen2019,Christensen2021}. Technical fluctuations of the prepared atom number do not impede our analysis and the statistical error on the obtained condensate fraction from the fit is smaller than the measured shot-to-shot fluctuations, which we therefore attribute to physical fluctuations of the BEC. The typical atom number at the peak of the fluctuations at $T/T_c^{0}=0.65$ is around 30,000 atoms. 

We model the density distribution as a continuous system with a modified interaction constant $g$ given by $g_{\rm eff}=g A_{\rm WS}/(2\pi\sigma^2)$ with the area of the Wigner Seitz cell $A_{\rm WS}$ and the harmonic oscillator length $\sigma$ on the lattice sites~\cite{Asteria2021}. In comparison to a continuous trap with the same scattering length, the interactions are enhanced by a factor $5$, which makes the semiideal model necessary and which is responsible for the larger shift of the fluctuation peak compared to ref.~\cite{Kristensen2019}

\textit{Theoretical Method.}
In order to theoretically describe the system, we employ the following Hamiltonian,

\begin{align}
      H=&-J \sum_{<i,j>,f} (\hat a^\dagger_{i,f} \hat a_{j,f}+h.c.)+\sum_{i,f} V_{i,f} ~ \hat n_{i,f} 
      \nonumber \\&
      +\frac{U}{2}\sum_{i,f} \hat n_{i,f} (\hat n_{i,f}-1),
      \label{H.eqn}
\end{align}
where $\hat a_{i,f}^\dagger$ and $\hat a_{i,f}$ are the creation and annihilation operators at site $i$ in mode $f$, $\hat n_{i,f} = a_{i,f}^\dagger a_{i,f}$ is the number operator, and $\langle i,j \rangle$ denotes nearest neighbor sites. Hamiltonian parameters are the single-particle tunneling $J$, the Bose-Hubbard interaction $U$, and the harmonic trapping potential, $V_{i,f} = V_{\textnormal{t}}(x_i^2 + y_i^2) + f \hbar \omega_z$. Here, $(x_i,y_i)$ are the spatial coordinates of site $i$ in the $xy$-plane, and the strength of the trap is given by $V_\textnormal{t} = m \omega^2 a_{\textnormal{lat}}^2/2=16.7~J$. 
The different modes $f$ are eigenmodes of the perpendicular trapping potential $V_{z,i}=  m \omega_z^2 z^2/2$. For the perpendicular trapping frequency, we choose  $\hbar\omega_z/J = 2.41$, to match the experimental conditions.

To analyze the system, we use the time-dependent Gross-Pitaevskii equation (GPE), which can be derived from the Heisenberg equation of motion applied to the second quantized Hamiltonian~\cite{Rev_Pitaev,Gies2004}.
Following the idea of the decomposition of the field operators into the classical field, describing the Bose condensation, and fluctuation part determining the quantum excitations, $ \hat a_{i,f}(t)\simeq (\Phi_{i,f}+\tilde{\psi}_{i,f}(t))e^{-i \epsilon_0 t/\hbar}
$
with $\Phi_{i,f}=\langle \hat a_{i,f}(t) \rangle$, beside applying the Hartree-Fock-Bogoliubov (HFB)  approximation and mean-field theory, we can derive the two coupled equations corresponding to the stationary GPE and the excitation particles. The derivation follows standard techniques as outlined in Refs.~\cite{Giorgini1996,Griffin_main,Suthar2015} and references therein:
\begin{align}
       \label{total_phi.eqn}
       &(V_{i,f}-\epsilon_0) \Phi_{i,f}-J \sum_{<j>}  \Phi_{j,f} 
        \nonumber\\ &
       +U (n^c_{i,f} \Phi_{i,f}+2\Phi_{i,f} \tilde{n}_{i,f}+\Phi_{i,f}^\dagger \tilde{\Delta}_{ii,f})=0,
       \\ & 
       \label{total_bogo.eqn}
     i \frac{d \tilde{\psi}_{i,f}}{dt}=-\sum_{<j>} J \tilde{\psi}_{j,f}+[ V_{i,f}-\epsilon_0 
    + 2U ( n_{i,f}^c+ \tilde{n}_{i,f})]\tilde{\psi}^\dagger_{i,f}
    \nonumber \\&
     +U (\Delta^{c*}_{ii,f}+\tilde{\Delta}^*_{ii,f}) \tilde{\psi}_{i,f}.  
\end{align}
Here, at each lattice site $i$, we define the number of particles in the condensate as $n^c_{i,f}=|\Phi_{i,f}|^2$, the condensate pairing as $\Delta^c_{ii,f}=\Phi_{i,f}^2$,  the number of non-condensate particles  as $\tilde{n}_{i,f}=\langle\tilde{\psi}^\dagger_{i,f} \tilde{\psi}_{i,f} \rangle$,
and  the pairing of non-condensed particles, as
$\tilde{\Delta}_{ii,f}=\langle\tilde{\psi}_{i,f} \tilde{\psi}_{i,f} \rangle$. The latter so-called anomalous terms are neglected (the Popov approximation) in the following.  The quantity $\epsilon_0$ stands for ground state energy of the condensate particles, and it is assumed that the total particle number $N=\sum_{i,f} \tilde{n}_{i,f}+n^c_{i,f}$
 remains fixed.  The two equations Eq.~\eqref{total_phi.eqn} and Eq.~\eqref{total_bogo.eqn} for the condensate and excited particles, respectively, are solved self-consistently yielding also the quasiparticle excitation spectrum and states.

\begin{figure}[t]
    \centering
\includegraphics[width=1\linewidth]{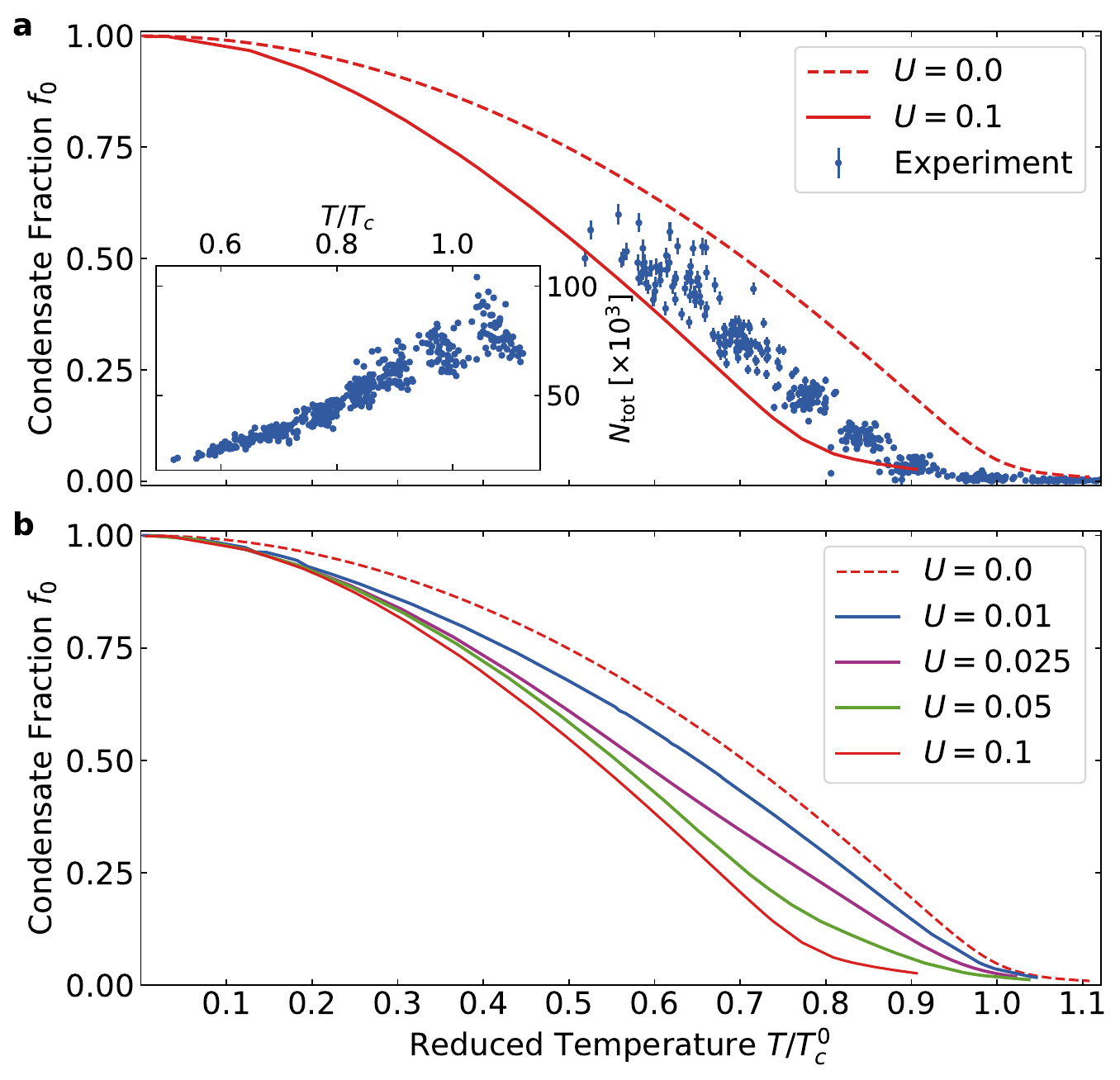}
    \caption{Condensate fraction across the phase transition for the interacting system. Numerical results for a fixed particle number $N = 1,000$ for the non interacting case (dashed red line) and the interacting case (solid red line) together with the experimental results (blue dots). Here, temperature is scaled by the non interacting temperature $T_c^{0}$ \cite{SupMat}. The experimental data show the values extracted from individual images and indicate strong fluctuations. The clustering of the data points is due to the finite steps of the final evaporation frequency in the preparation.
    The inset shows the influence of varying Hubbard interaction strength $U/J = 0.0, 0.01, 0.025, 0.05, 0.1$ in the numerics. }
    \label{fig:condensation_Utune}
\end{figure}

We thus obtain the condensation fraction and analyze it below  as function of the number of particles in the system.
In order to study fluctuations of the condensate, one needs to obtain the fluctuations of the  non condensed particles from which the ground state fluctuations can be inferred. However, it is known that the Bogoliubov-Popov approximation breaks down at intermediate temperatures (close to the critical temperature).
Since we are interested in the description of a system with fixed particle number and over a wide range of temperatures, we employ therefore a hybrid method combining the Bogoliubov approach with the master equation method introduced in \cite{Kocharovsky2006,Scully_2006,scully_2010,scully_PRL} that was shown to yield well behaved predictions for condensate fluctuations even across $T_c$.  In this approach, the non-condensed particles act as a reservoir that facilitates the thermalization of the condensate (by exciting particles from the condensate through the absorption of Bogoliubov quasi-particles and vice versa, see Fig.~\ref{fig:system}a).  A master equation for the condensate is then obtained which  depends on the statistical properties of the Bogoliubov quasiparticles which are available from the solution of Eqn.~(\ref{total_bogo.eqn}). The master equation technique introduces thermal corrections to the condensate fluctuations (see Sec.~2.4 of Supp. Mat.~\cite{SupMat}).

\textit{Results.}
We start by comparing the condensate fraction as a function of reduced temperature of the experiment and the numerics for $N=1,000$ particles (Fig.~2). Both curves show a smooth function resulting from the finite system size. We repeat the numerics for different interaction strength between $U=0$ and $U=0.1 J$ and find an expected shift of the critical temperature to smaller values \cite{Giorgini1996}. The range of interaction strengths fits with an estimate of the effective Hubbard interactions in the experiment \cite{Zahn2022}. 

Next, we compare the condensate fluctuations as a function of reduced temperature (Fig.~3). We find curves with a peak of the fluctuations at a temperature that shifts with interaction strength and reaches approximately $T \sim 0.65 T^{0}_{c}$ for the interaction strength $U/J=0.1$. In the numerics, we determine the scaling with atom number by varying the atom number while keeping all other parameters fixed and fitting $\delta N_{\rm BEC}^2 \propto N^{1+\gamma}$. We find strongly anomalous scaling with an exponent $\gamma_{\rm theo}= 0.74$, which is approximately independent of temperature
in the absence of interactions. The same value holds for interacting systems irrespective of the interaction strength at sufficiently large temperature, whereas strong finite-size deviations are observed at low temperature, see also Sec.~2.3 of Supp. Mat. \cite{SupMat}.

In the experiment, the temperature and atom number are correlated, because we vary the temperature by different end points of the evaporation ramp. We therefore extract the exponent of the anomalous fluctuations by considering the overall shape of the fluctuation curve as a function of the temperature normalized to the atom-number dependent critical temperature of the non-interacting system. We compare the curve of the theoretical fluctuation data normalized by $1/N^{1+\gamma_{\rm theo}}$ with the experimental fluctuation data normalized by $1/N^{1+\gamma_{\rm exp}}$ and find a good agreement for $\gamma_{\rm exp}=0.62$, i.e., a value with a mismatch of less than 20\% between $\gamma_{\rm theo}$ and $\gamma_{\rm exp}$. This gives strong evidence to establish these strongly anomalous fluctuations for our 2D/3D crossover geometry. We note that the exponents are in between the expected values for non-interacting systems in a truly 2D lattice ($\gamma=1$) and a truly 3D lattice ($\gamma=1/3$). 

\begin{figure}
    \centering
    \includegraphics[width=1
    \linewidth]{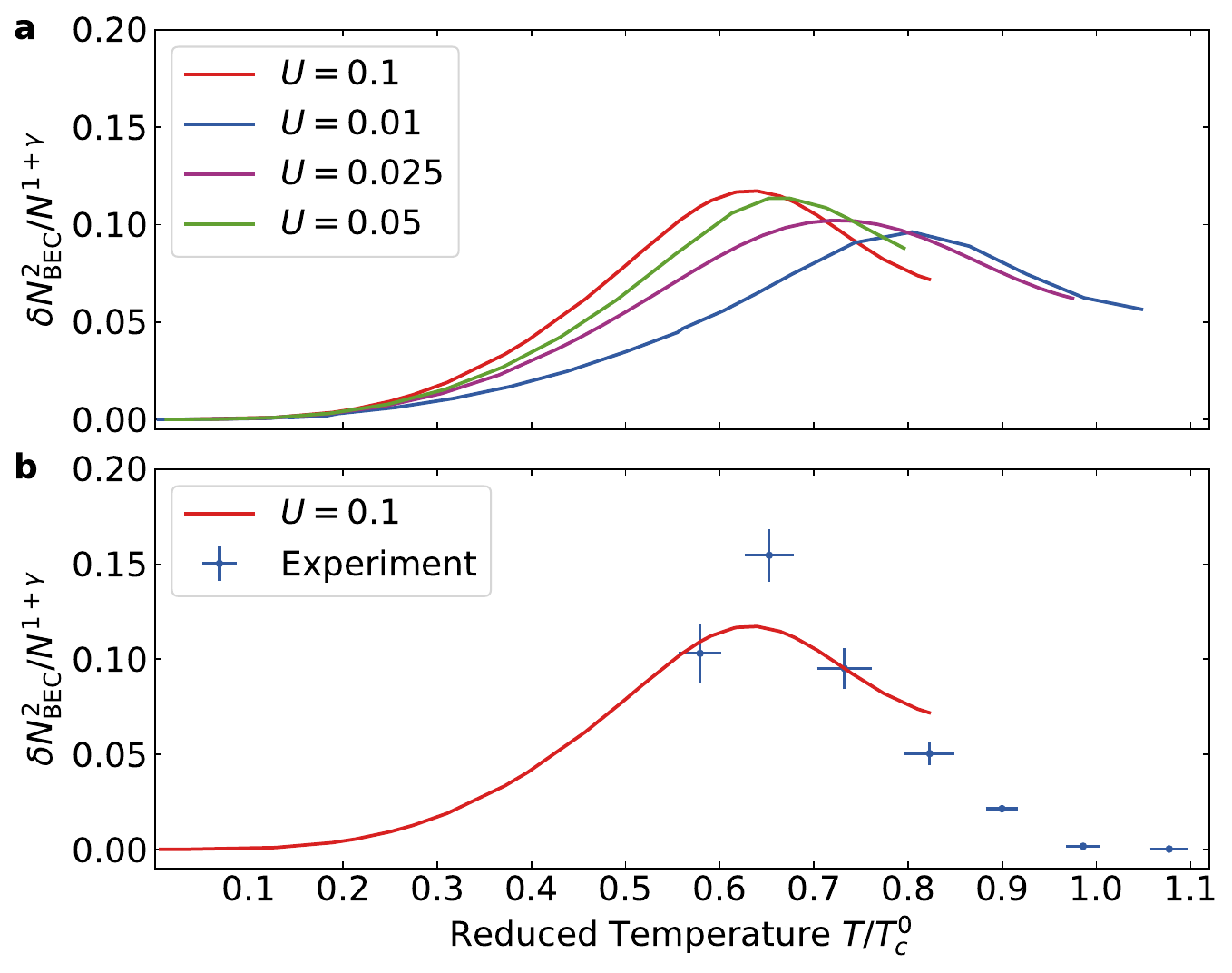}
    \caption{Condensate fluctuation across the phase transition. (a) Variance of the condensate particle number from the numerics normalized by the total atom number $N=1,000$ with the anomalous scaling exponent $\gamma_{\rm theo} = 0.74$. When the Hubbard interaction strength is increased from $U/J=0.01$ to $U/J=0.1$, the peak of the fluctuations shifts from $T/T_c^0=0.8$ to $T/T_c^0=0.65$. (b) Variance of the condensate particle number in the experimental data (blue points) together with the numerics for $U/J=0.1$ (solid line). The data agrees well for an experimental scaling exponent $\gamma_{\rm exp}=0.62$, which is close to the theoretical value. The theory curves do not bend down at high temperatures, because the approach becomes less reliable close to the phase transition.}
    \label{fig:Fluctuation_TE}
\end{figure}
    
\textit{Conclusions.}
In conclusion, we have studied the condensate fluctuations of an interacting BEC in a triangular lattice of tubes both experimentally and theoretically and found strong anomalous scaling with the atom number with an exponent of $\gamma_{\rm theo}=0.74$ and $\gamma_{\rm exp}=0.62$. Our finding of anomalous fluctuations in large systems sizes of 30,000 atoms is an important contribution to the debate of anomalous scaling in the thermodynamic limit \cite{Kocharovsky2006,Kruk2025}.
In the future, it would be interesting to study the BEC fluctuations in an atomic cloud, which thermalizes with a heat bath of a second species or where losses are introduced \cite{Paulino2024} or for BECs in higher bands of an optical lattice \cite{Wirth2011}. The study could be extended to full counting statistics \cite{Rath2010}, e.g., after an interaction quench \cite{Horvath2024}. Moreover, local fluctuations \cite{Astrakharchik2007} or higher moments of the distributions \cite{Bureik2024} also provide valuable insights into BECs. The role of interactions on the condensate fluctuations is not fully understood and more studies are required, also including dipolar interactions. A detailed understanding of condensate fluctuations will become important in quantum metrology, e.g., for producing entangled atom pairs in atom interferometers via density dependent processes \cite{Anders2021}.

{\bf Data Availability Statement}
The data that support the plots presented in this paper and other findings of this study are available from the corresponding author upon reasonable request. The authors declare no competing financial interests.

{\bf Acknowledgements}
We thank Marcel Kosch, Henrik Zahn and Moritz Sträter for contributions in an early stage of the project. This work was funded by Deutsche Forschungsgemeinschaft (DFG, German Research Foundation) via Research Unit FOR 5688, Project No. 521530974, and via the cluster of excellence AIM, EXC 2056—project ID 390715994 as well as by ’Hamburg Quantum Computing’, financed by the city of Hamburg and the European Union. U.B. acknowledges financial support of the IBM Quantum Researcher Program. R.W.C. acknowledges support from the Polish National Science Centre (NCN) under Maestro Grant No. DEC- 2019/34/A/ST2/00081. 
T.G. acknowledges funding by the Department of Education of the Basque Government through the IKUR Strategy, through BasQ (project EMISGALA), and through PIBA\_2023\_1\_0021 (TENINT), as well as by the Agencia Estatal de Investigación (AEI) through Proyectos de Generación de Conocimiento PID2022-142308NA-I00 (EXQUSMI), and that this work has been produced with the support of a 2023 Leonardo Grant for Researchers in Physics, BBVA Foundation. The BBVA Foundation is not responsible for the opinions, comments, and contents included in the project and/or the results derived therefrom, which are the total and absolute responsibility of the authors.
ICFO-QOT group acknowledges support from:
European Research Council AdG NOQIA;
MCIN/AEI (PGC2018-0910.13039/501100011033,  CEX2019-000910-
S/10.13039/501100011033, Plan National FIDEUA PID2019-106901GB-I00, Plan National
STAMEENA PID2022-139099NB, I00, project funded by
MCIN/AEI/10.13039/501100011033 and by the “European Union
NextGenerationEU/PRTR $\&$ quot; (PRTR-C17.I1), FPI); QUANTERA DYNAMITE PCI2022-
132919, QuantERA II Programme co-funded by European Union’s Horizon 2020 program
under Grant Agreement No 101017733; Ministry for Digital Transformation and of Civil
Service of the Spanish Government through the QUANTUM ENIA project call - Quantum
Spain project, and by the European Union through the Recovery, Transformation and
Resilience Plan - NextGenerationEU within the framework of the Digital Spain 2026
Agenda; MICIU/AEI/10.13039/501100011033 and EU (PCI2025-163167);
Fundació Cellex;
Fundació Mir-Puig;
Generalitat de Catalunya (European Social Fund FEDER and CERCA program;
Barcelona Supercomputing Center MareNostrum (FI-2023-3-0024);
Funded by the European Union. Views and opinions expressed are however those of the
author(s) only and do not necessarily reflect those of the European Union, European
Commission, European Climate, Infrastructure and Environment Executive Agency
(CINEA), or any other granting authority. Neither the European Union nor any granting
authority can be held responsible for them (HORIZON-CL4-2022-QUANTUM-02-SGA
PASQuanS2.1, 101113690, EU Horizon 2020 FET-OPEN OPTOlogic, Grant No 899794,
QU-ATTO, 101168628), EU Horizon Europe Program (This project has received funding
from the European Union’s Horizon Europe research and innovation program under grant
agreement No 101080086 NeQSTGrant Agreement 101080086 — NeQST);
ICFO Internal “QuantumGaudi” project;


%

\newpage
\appendix
{\bf Supplementary Material}
\section{1. Experimental Data Pipeline}

\subsection{1.1. Calculation of the Site Occupation}
To determine the length and rotation of the lattice vectors we fit the peaks of the Fourier transformation of the optical density. The resulting lattice vector length is $5.36 px$ and the lattice rotation $-0.45°$. For each individual measurement, we calculate the lattice phase by a three step algorithm. First the images are oversampled by a factor of 10 to allow integration masks with features below pixel resolution. Secondly we find the central peak of the lattice by a star shape evolution maximum criterion. Using the found lattice vectors we then maximize the integrated signal of a lattice with hexagonal cells shrinked to 1/3 of the expected size by varying the lattice phase. The lattice site occupation is calculated using a mask of Wigner-Seitz cells and summing up the overall signal inside the cell. The resulting integrated optical densities are translated into particle numbers using the factor $(a_\textnormal{pixel} / m_\textnormal{optical})^2 \sigma_0$ with the pixel size of the camera $a_\textnormal{pixel} = 13 \mu m$, the optical magnification $m_\textnormal{optical} = 2$ and the scattering cross section $\sigma_0 = 0.2907 \mu m^2$.

\subsection{1.2 Fitting the Density Distribution}
For modeling the density distribution, we follow our previous work \cite{Asteria2021}. The 3D density of the BEC can be described according to the Thomas-Fermi approximation by
\begin{equation}
    n_\mathrm{BEC} (x,y,z) = \frac{15}{8 \pi} \frac{N_\mathrm{BEC}}{R_\rho^2 R_z} \mathrm{max} \left( 1 - \frac{\rho^2(x,y)}{R_\rho^2} - \frac{z^2}{R_z^2}, 0\right)
\end{equation}
which can be reformulated using the relations $R_\rho = c N_\mathrm{BEC}^{\frac{1}{5}}$ and $R_z = R_\rho \frac{\omega_\rho}{\omega_z}$ to a density relying on the number of atoms inside the BEC $N_\mathrm{BEC}$ and the factor $c = 0.352 \,\mathrm{\mu m}$. $c$ can be obtained by
\begin{equation}
    c = \left(15 \frac{g_\textnormal{eff}}{g} \frac{a_\textnormal{sc}}{\bar a}\right)^\frac{1}{5} \sqrt{\frac{\hbar \bar\omega}{m \omega_\rho^2}}
\end{equation}
according to \cite{Pethick2008} using $\bar\omega = \left(\omega_\rho^2 \omega_z \right)^{\frac{1}{3}}$ and $\bar a = \sqrt{\frac{\hbar}{m\bar\omega}}$ using a scattering length of $a_\textnormal{sc} \approx 100 \,\textnormal{Bohr}$ leading to the interaction constant of $g = 4\pi \frac{\hbar^2 a_\textnormal{sc}}{m}$. The lattice nature of the system can be accounted by a renormalization of the interaction constant $g_\textnormal{eff} = g \frac{A_\textnormal{WS}}{2\pi \sigma^2}$ with $A_\textnormal{WS}$ is the size of the Wigner-Seitz cell and $\sigma = \sqrt{\frac{\hbar}{m\omega_\textnormal{onsite}}}$ the radial oscillator length. The onsite trap frequency can be calculated from the lattice depth given by $\hbar \omega_\textnormal{onsite}=3 E_\textnormal{rec} \sqrt{\frac{2V_\textnormal{lat}}{E_\textnormal{rec}}}$. For the relevant dataset the lattice depth is $V_\textnormal{lat} = 1 E_\textnormal{rec}$.

Using a semi-ideal approach, the thermal density can be described by an ideal gas in a potential given by a superposition $V(x,y,z) = V_\textnormal{trap} (x,y,z) + V_\textnormal{BEC}(x,y,z)$ of the harmonic  external trap $V_\textnormal{trap}(x,y,z)$ defined by $\omega_\rho$ and $\omega_z$ and the repulsion term from the condensed atoms $V_\textnormal{BEC}(x,y,z) = 2g_\textnormal{eff} n_\textnormal{BEC}(x,y,z)$. The density of an ideal Bose gas in semi-classical approximation is given by
\begin{equation}
    n_\textnormal{th} = \frac{\textnormal{Li}_\frac{3}{2}\left(e^{-\beta (V(x,y,z) - \mu)}\right)}{\lambda_T^3}
\end{equation}
with the polylogarithm $\textnormal{Li}_s(x)=\sum_{k=1}^\infty \frac{x^k}{k^s}$ and $\lambda_T = \hbar \sqrt{\frac{2\pi}{mk_BT}}$.

We use the combined integrated density function
\begin{equation}\label{eq:bimodal}
    n(x, y) = \int \textnormal{d}z n_\textnormal{th}(x,y) + n_\textnormal{BEC}(x,y)
\end{equation}
as fitting function as shown in figure \ref{fig:system}c. The free fit parameters are the center of the distribution $\hat x, \hat y$, the temperature $T$, the chemical potential $\mu$, the number of particles inside the BEC phase $N_\textnormal{BEC}$ and a global offset. The fitting routine is implemented using a two step routine by first fitting a Gaussian distribution to find good starting parameters and second step where we use the full function. The integration is numerically implemented by the a 1,000 step Riemann integration with a maximum value of the integration value $z_\textnormal{max} = 0.0005\ m$. As the fit function does not include all physical effects of the underlying physical system, not all measurements are well described by the fit function. We remove all results from the dataset where the difference in the overall particle number is larger than 1,000 between the summed up onsite occupations and the total particle number obtained by the fit, the overall particle number is less than 20,000 ot the R-squared fit value is lower than 0.99. This reduces the number of data points from 1,355 to 432. 

\subsection{1.3 Derivation of the critical temperature}
As the measurement methods we use in the experiment do not ensure a fixed total atom number it is necessary to derive the critical temperature $T_c$ for each data point individually using theoretical considerations of the density of states. The details of the derivation can be found in the methods section of \cite{Asteria2021}. An appropriate approximation for the number of states is given by
\begin{equation}
\label{eq:number_of_states}
    N(E) = \left( \frac{E}{E_0} \right)^2 + \max \left(\frac{1}{6} \left( \frac{E - \hbar \Delta_g}{\hbar \bar{\omega}} \right)^3, 0 \right)
\end{equation}
where $E_0 = \sqrt{\frac{\hbar A_\textnormal{WS} m \bar\omega^3}{\pi}} = h \times 57 \,\textnormal{Hz}$ and the energy of the bandgap $\Delta_g = 2\pi \times 6040 \textnormal{Hz}$ which is obtained by theoretical band structure calculations without an external trapping potential. Using equation \ref{eq:number_of_states} the density of states reads as \begin{equation}
    g(E) = \frac{\textnormal{d}N}{\textnormal{d}E} = \frac{2E}{E_0^2} + \left\{
    \begin{array}{ll}
    \frac{\left(E - \hbar \Delta_g\right)^2}{2 \hbar \bar\omega} & E \geq \Delta_g \\
    0 & E < \Delta_g
    \end{array}
    \right.
\end{equation}
which allows to calculate the number of excited atoms numerically by solving the integral
\begin{equation}
    N_\textnormal{exc} = \int \textnormal{d}E \frac{g(E)}{e^{\frac{E}{k_\textnormal{B}T}} - 1}.
\end{equation}
As the number of particles inside the BEC $N_\textnormal{BEC}$ and the total particle number $N_\textnormal{tot}$ are determined by the fit, the critical temperature can be determined by solving
\begin{equation}
    N_\textnormal{tot} = N_\textnormal{exc}(T_c)
\end{equation}
for each data point.

\subsection{1.4 Fluctuations}
To calculate the fluctuations we bin the data in 7 equidistant bins between the minimal value of $\min(T/T_c) = 0.52$ and the maximum value $\max(T/T_c) = 1.12$. For each bin we fit a linear function to the $N_\textnormal{BEC}$ vs. $T/T_c$ data and calculate the average of the distance of each point in $N_\textnormal{BEC}$ direction to the linear approximation. The error of the variance is the standard deviation normalized by the square root of the number of measurements in the bin. The $T/T_c$ value for each bin is the averaged value of the calculated $T/T_c$ as described in the previous section.

\section{2. Theoretical Considerations}
\subsection{2.1 Non-interacting BEC in optical lattice}
\begin{figure}[b]
    \centering    
    \includegraphics[width=1\linewidth]{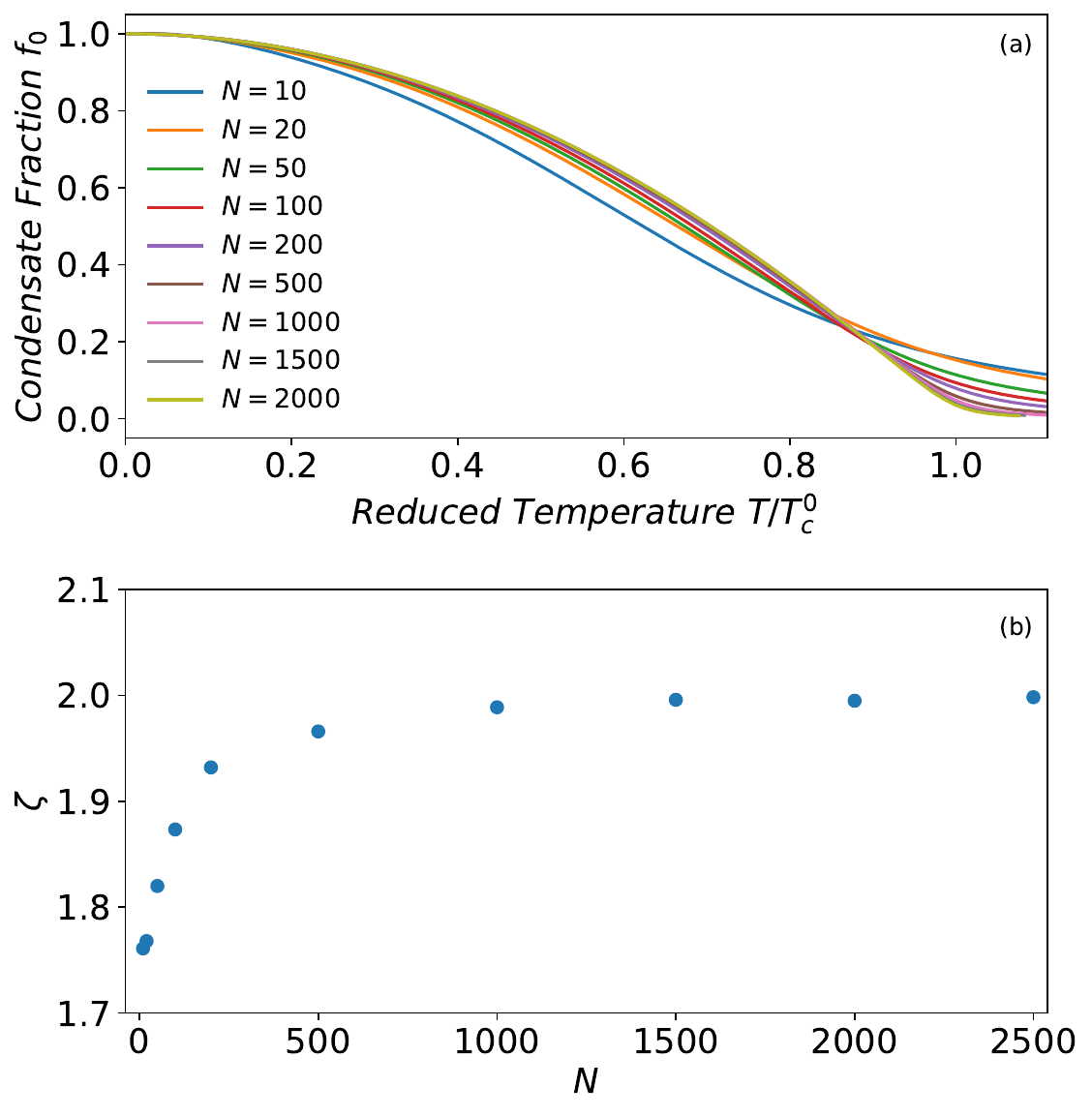}
    \caption{Bose gas in a triangular optical lattice in the absence of Bose-Hubbard interaction $U=0$. Panel~(a) illustrates the condensation fraction as a function of temperature for various particle numbers, with temperature scaled by the critical temperature $T^{0}_{c}$, which is determined through a power-law fit. Panel~(b) shows the power-law description of the condensation fraction using the power-law relation $N_0/N = 1 - (T/T^{0}_{c})^\zeta$, demonstrating the dependence on different particle numbers $N$. The perpendicular frequency $\omega_z/J = 2.41$ and a fixed in-plane trap potential $V_\textnormal{trap}/J = 16.7$ are assumed throughout the analysis.
    }
\label{fig:N_Vt=16.7_z=2.41_U=0}
\end{figure}

In this section, we investigate the properties of a non-interacting BEC $U=0$ confined within a triangular optical lattice. 
To explore the behavior of the BEC under these conditions, we analyze the condensation fraction as a function of the number of particles in the system.
In this regard, one should solve the two self-consistent equations Eq.~\eqref{total_phi.eqn} and Eq.~\eqref{total_bogo.eqn}~\cite{Suthar2015, Pethick2008} for the condensate and excited particles, respectively.
The results in Fig.~\ref{fig:N_Vt=16.7_z=2.41_U=0}~(a) shows the condensation fraction for different numbers of particles. This analysis provides insight into the distribution of particles within the lowest energy state, revealing how the system approaches macroscopic occupation of the ground state as the number of particles increases. 
We focus on the condensation fraction $N_0/N$ as a function of temperature, with the temperature scaled by the critical temperature $T^{0}_{c}$, obtained from fitting the data to a power-law function. As shown in Fig.~\ref{fig:N_Vt=16.7_z=2.41_U=0}~(a), as the temperature ($T/T^{0}_{c}$) increases, the particles gradually leave the condensate and transition to excited states due to thermal fluctuations.  It's important to note that the temperature in this figure is rescaled by $T^{0}_{c}$, and as the number of particles (or particle density) increases, the critical temperature $T^{0}_{c}$ also increases. 
 We quantify the power-law behavior of each condensation fraction curve in Fig.~\ref{fig:N_Vt=16.7_z=2.41_U=0}~(b), in which $y$-axis represent power-law exponent $\zeta$ in terms of particle number $N$. 
 The exponent $\zeta$, reveals that as the number of particles increases, $\zeta$ also increases, and for a number of particles high enough it approaches to the $\zeta \sim 2$.
 Comparison of this result with the continuous anisotropic Bose gas (in which $\zeta \sim 2.77$), indicates that the presence of the optical lattice suppresses the exponent $\zeta$. 
 This suppression is primarily due to the kinetic energy term in the single-particle tight-binding Hamiltonian, which is influenced by the optical lattice. Hence, this reduction in $\zeta$ indicates that the system's behavior is more characteristic of a 2D structure with continuous trap in which  $\zeta=2$.

\subsection{2.2 Interacting BEC in optical lattice}
\begin{figure}[t]
    \centering
\includegraphics[width=1\linewidth]{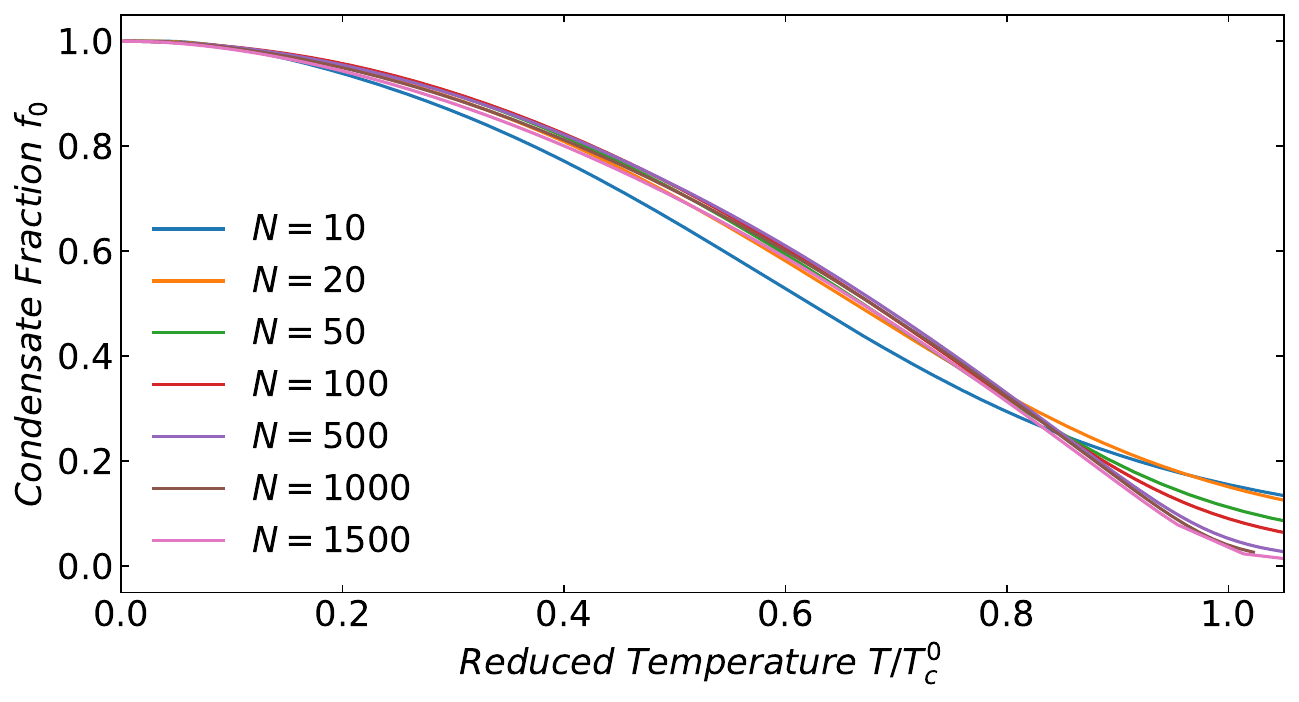}
    \caption{Condensation fraction of the trapped Bose gas in a triangular lattice in the presence of Bose-Hubbard interaction in terms of different particle numbers. Here, $T$ is scaled by non interacting temperature $T_c^{0}$, for different particle numbers with a fixed Bose-Hubbard interaction $U/J=0.005$. All the rest of parameters are the same as those in Fig.~\ref{fig:N_Vt=16.7_z=2.41_U=0}.}
    \label{fig:N_Vt=16.7_z=2.41_U=0.005}
\end{figure}

In this section, we explore the impact of the Bose-Hubbard interaction on the condensation fraction of the trapped BEC. In the extreme limit, the Bose-Hubbard interaction tends to drive the system towards a Mott insulator phase, where particles become localized and condensation is suppressed. To understand the role of interaction, we begin by analyzing the effect of a very small Bose-Hubbard interaction on the condensation fraction across different particle numbers, as depicted in Fig.~\ref{fig:N_Vt=16.7_z=2.41_U=0.005}.
The results reveal two distinct behaviors as the particle number increases. At lower particle numbers, $N\le 500$, the introduction of the Bose-Hubbard interaction does not show major changes in the behavior of the condensed fraction compared to the case without interaction. 
However, as the number of particles increases, the system is depleted rapidly, seen as a downward shift of the condensate fraction curve.
This behavior is fundamentally related to the nature of the Bose-Hubbard interaction, which is proportional to $|\Phi|^2$, where $ \Phi$ denotes the condensate wave function. As the particle number $N$ increases, the interaction strength $U$, which scales with the square of the condensate density, becomes more significant. The stronger interaction leads to more pronounced effects on the system's coherence, resulting in a faster depletion of the condensate. This interaction-induced depletion is reflected in the reduction of the exponent $\zeta$. Namely, in the non-interacting case we reached to the $\zeta=2$ for the high number of particles while the inclusion of interaction in Fig.~\ref{fig:N_Vt=16.7_z=2.41_U=0.005} results in the smaller exponent $\zeta\simeq 1.67$ for $N=2000$.

\subsection{2.3 Atom Number Fluctuations in Optical Lattice Bose Gas}
In this section, we investigate the BEC atom number variance/ fluctuations by examining the anomalous scaling of fluctuations in both non-interacting and interacting Bose gases confined in an optical lattice. The calculation of the condensate fluctuations within the grand canonical ensemble leads to an unphysical divergent result at the critical temperature $T = T_c$, which contradicts experimental observations. Conversely, when the microcanonical or canonical ensembles are employed, this divergence does not appear, even in the thermodynamic limit~\cite{Christensen2021,Weiss1997}.

The divergence in the grand canonical ensemble arises due to the ensemble's inherent flexibility in particle number, which leads to the exaggerated fluctuations near the critical temperature. This issue is particularly relevant in systems where the number of particles is fixed, such as in experiments, making the microcanonical and canonical ensembles more accurate representations. In these ensembles, the constraints on particle number help regulate fluctuations, yielding results that align more closely with experimental data.

Solving the coupled HFB Hamiltonian Eq.~\eqref{total_bogo.eqn} we can calculate the atom number fluctuations for the non-condensate particles using
\begin{equation}
    \Delta \tilde{N}^2=\sum_{i,f} \delta\tilde{n}_{i,f}^{2}=\sum_{i,f} \langle\Tilde{\psi}^\dagger_{i,f} \Tilde{\psi}_{i,f} \Tilde{\psi}^\dagger_{i,f}\Tilde{\psi}_{i,f}\rangle-\langle\Tilde{\psi}^\dagger_{i,f} \Tilde{\psi}_{i,f}\rangle^2,
\end{equation}
which gives rise to
\begin{align}
    \label{tildeflx.eq}
    &\delta\tilde{n}_{i,f}^2=
    \\ &
    \sum_{b,p}
    u^*_{ib,f}v_{ib,f} u_{ip,f}v^*_{ip,f}[2N(E_{b,f})+1][2N(E_{p,f})+1]
    \nonumber\\&+[(|u_{ib,f}|^2+|v_{ib,f}^2|)N(E_{b,f})+|v_{ib,f}^2|]
     \nonumber\\&
     \times [(|u_{ip,f}|^2+|v_{ip,f}^2|)N(E_{p,f})+|u_{ip,f}^2|],
\end{align}
Here, $N(E_{b,f})$ is the Bose-Eistein distribution at energy $E_{b,f}$, summation $b ,p $ stands for summation on different Bogoliubov bands. Considering that the atom number fluctuations are equivalent to those of the non-condensed particles~\cite{Meier1999}, we again encounter the unphysical at higher temperatures originated from the break down of the Popov approximation at higher temperature $T>T_c$. In our numerical simulation, although the divergent behavior in the atom number fluctuation would not occur, we used a hybrid approach  with the master equation techniques developed by Scully et al. (see  Supp. Mat. Sec. 2.4 \cite{SupMat} for details) to obtain more reliable results at higher temperatures.

\begin{figure}[t]
    \centering
\includegraphics[width=1\linewidth]{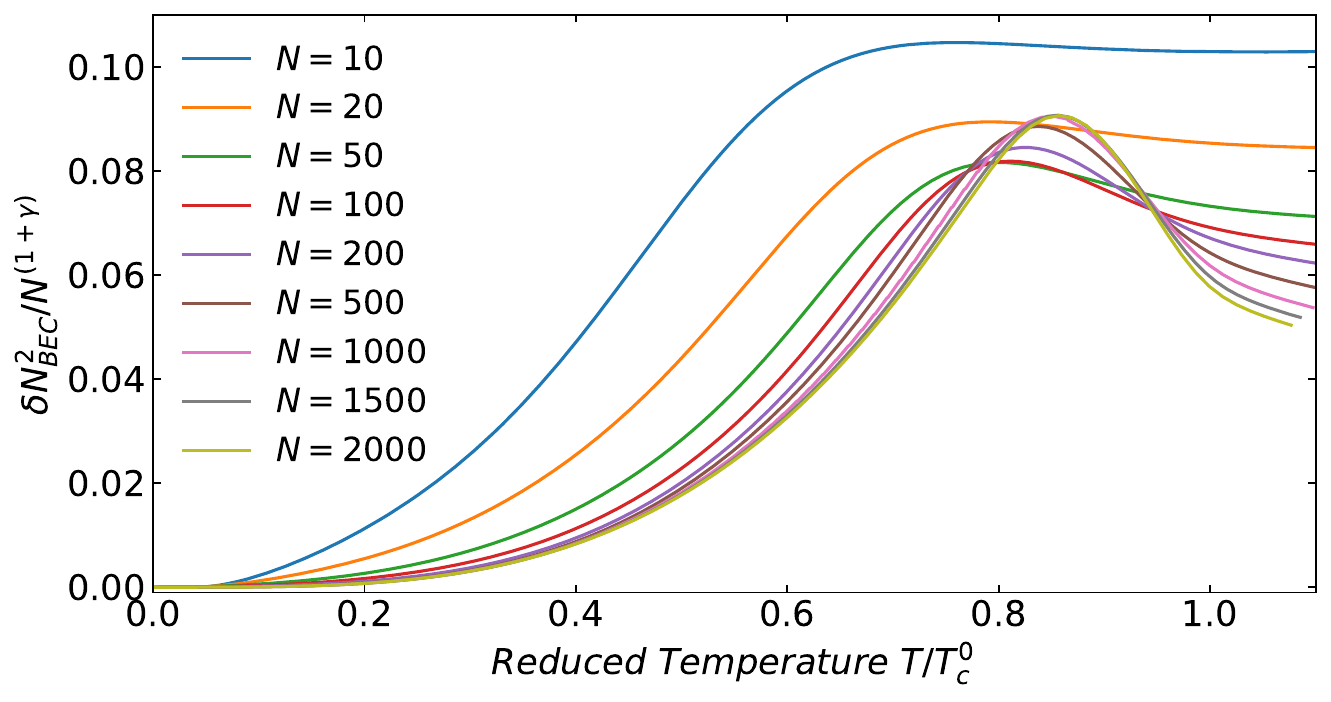}
    \caption{Condensate Fluctuations of the trapped non-interacting Bose gas in a triangular lattice for various particle numbers. 
    The atom number fluctuations are scaled with particle number $N^{1+\gamma}$ and the scaling parameter is $\gamma=0.74$. 
    All other parameters are identical to those in Fig.~\ref{fig:N_Vt=16.7_z=2.41_U=0}.
    }
    \label{fig:flx_nI_lattice}
\end{figure}

We begin by examining fluctuations in a non-interacting optical Bose gas for various particle numbers in Fig.~\ref{fig:flx_nI_lattice}. Here, the atom number fluctuations are scaled with particle number $N^{1+\gamma}$ in which  in order to determine the exponent $\gamma$, we fit the peak of the atom number fluctuations with the particle number to minimizes the variance of all peak values particularly at higher particle numbers.

The asymptotic behavior observed across all curves,  resembles the same properties, namely, increasing the total particle number, enhances the fluctuations and result in a pronounced peak at temperatures near the critical temperature $T_c$. To the right of this peak, the fluctuations decrease at higher temperatures.
Here, the scaling of the atom number fluctuations with $ N^{1+\gamma}$ reveals the presence of anomalous fluctuations, with a scaling exponent $\gamma = 0.74$. 

\begin{figure}[t]
    \centering
\includegraphics[width=1\linewidth]{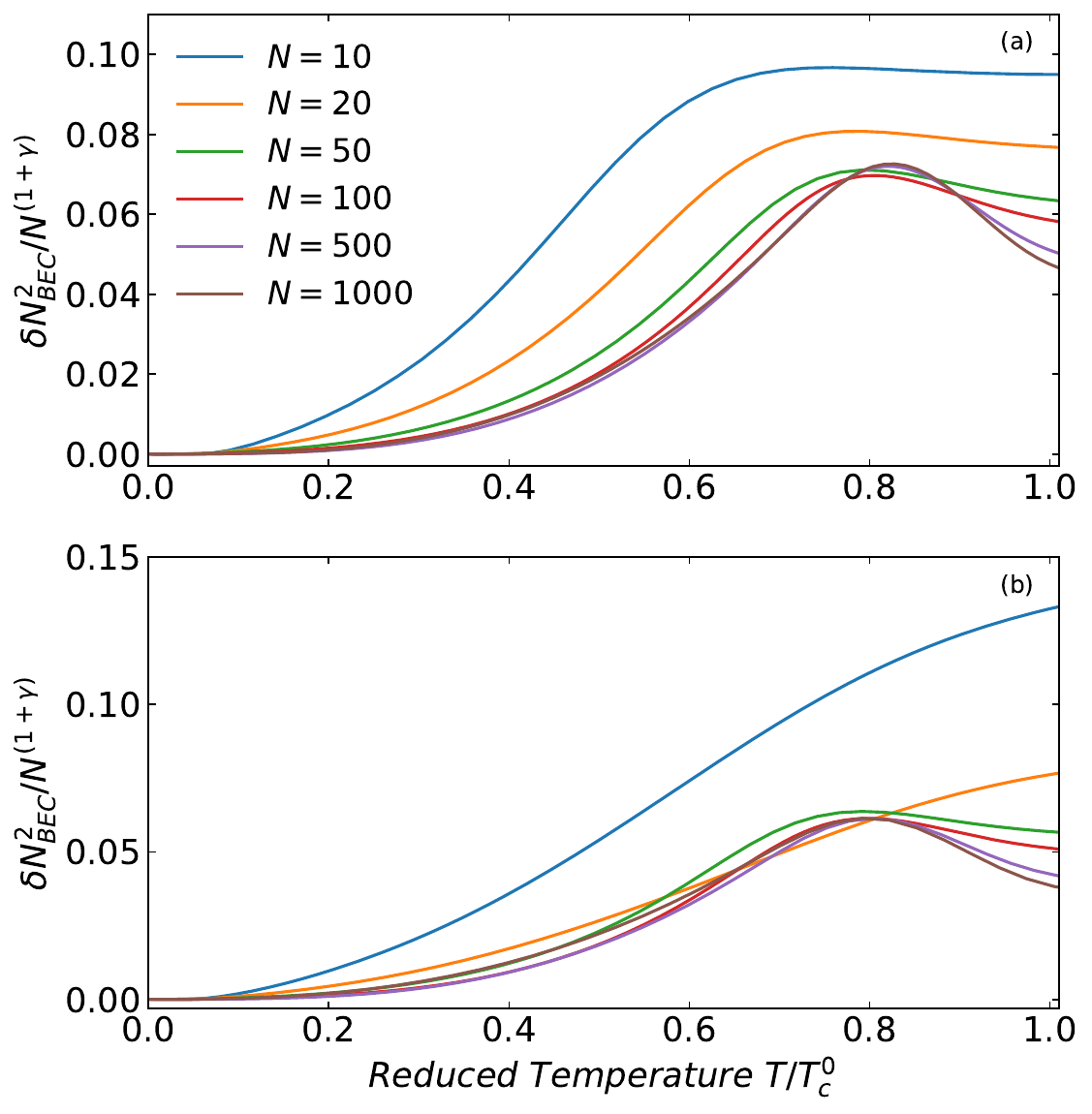}
    \caption{Condensate Fluctuation of the trapped  Bose gas in triangular lattice in the presence of Bose Hubbard interaction $U/J=0.005,0.01$ in terms of different number of particles in panel~(a) and (b), respectively. The anomalous scaling in panel~(a) and (b) is $\gamma\sim 0.77$ and $0.80$, respectively.
    The rest of parameters is same as Fig.~\ref{fig:N_Vt=16.7_z=2.41_U=0}.}
    \label{fig:flx_U}
\end{figure}

The remaining question to address is how interactions influence the condensate fluctuations, particularly when considering different particle numbers $N$ and varying strengths of the Bose-Hubbard interaction $U$. To this end, we examine the behavior of condensate fluctuations under specific interaction strengths, with $U/J= 0.005$ and $U/J = 0.01$ shown in Fig.~\ref{fig:flx_U} in panels~(a) and~(b), respectively.
The results allow us to observe how the interplay between particle number and interaction strength modifies the fluctuation behavior. As the interaction strength $U$ increases, we expect to see a shift in the fluctuation, particularly at critical points, which could indicate a suppression or enhancement of fluctuations depending on the balance between kinetic energy and interaction energy. Additionally, varying the particle number $N$ provides insights into finite-size effects and how they interplay with interactions in determining the overall fluctuation characteristics.

The results in the Fig.~\ref{fig:flx_U} indicate the anomalous exponent of the fluctuations, obtained through scaling with $N^{(1+\gamma)}$ is $\gamma \sim 0.77$  for $U/J=0.005$. However, as the interaction strength increases, the exponent rises to $\gamma \sim 0.80$ for $U/J=0.01$. This trend demonstrates that increasing the interaction strength leads to greater depletion, which in turn enhances the fluctuations. This enhancement is reflected in the larger anomalous scaling exponent $\gamma$ observed at higher interaction strengths, highlighting the significant role that interactions play in driving fluctuations within the condensate.

Furthermore, examining the temperature at which the peak of the fluctuations occurs reveals another important effect: as the interaction strength increases, the peak shifts to a lower temperature. This shift is consistent with the interaction-induced depletion of particles, which reduces the temperature at which the maximum fluctuations are observed. 

\begin{figure}[t]
    \centering
\includegraphics[width=1\linewidth]{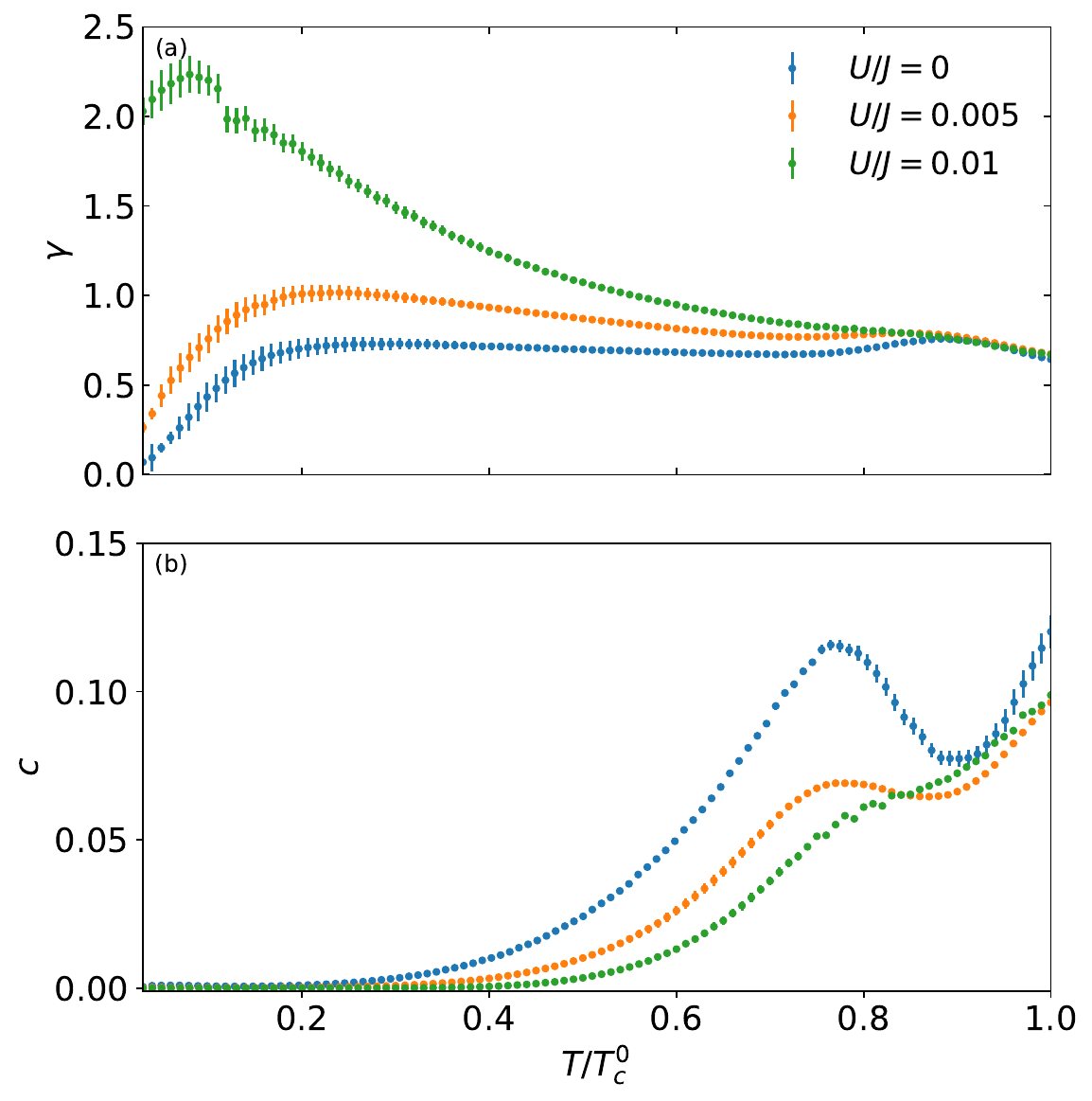}
    \caption{Anomalous exponent of fluctuation.
    The anomalous exponent $\gamma$ is extracted by fitting the condensate fluctuation data at each temperature to the function $c N^{(1+\gamma)}$, where $N$ is the total particle number. Panel~(a) and (b) indicates exponent $\gamma$ and coefficient $c$, respectively. This fitting yields a temperature-dependent exponent that varies smoothly, particularly near the peak of the fluctuations. The plots show the results for different interaction strengths: $U/J=0,0.005,0.01$.  }
    \label{fig:gamma}
\end{figure}

In this analysis, the anomalous exponent is determined by fitting the condensate fluctuation data calculated for various total particle numbers to the functional form $c N^{(1+\gamma)}$ at each temperature. 
The results of the fitting in terms of different temperature are shown in Fig.~\ref{fig:gamma}, for three different interaction strengths: $U/J=0,0.005,0.01$. here, panel~(a) shows the anomalous exponent $\gamma$ and fitting coefficient $c$ is shown in panel~(b). This fitting procedure yields a distinct value of the exponent $\gamma$ for every temperature point and interaction strength. We find that the interacting and non-interacting $\gamma$ totally agree for $T>0.85T_c$ and are almost temperature-independent. At smaller $T$, the interacting $\gamma$ develops temperature-dependencies which we attribute to the finite system size. The values of $\gamma$ at the respective peak of the fluctuations are $\sim 0.74$, $\sim 0.77$, $\sim 0.80$, and $\sim 0.82$ for interaction strengths of $U/J=0$, $U/J=0.005$, $U/J=0.01$ and $U/J=0.1$ (not shown). For the latter interaction strength, the calculated exponent is based on less extensive numerics than the rest. These observations justify the use of a single constant exponent $\gamma=0.74$ in Fig.~\ref{fig:flx_nI_lattice} and~\ref{fig:flx_U} and for the comparison with experiment in the relevant temperature range.

\subsection{2.4 Master Equation Method}
In order to calculate the condensate fluctuations of the BEC gas, we used a hybrid approach. This approach is based on calculation of condensate particles and atom number fluctuation from Bogoliubov Hamiltonian and incorporating the correction terms originated from master equation technique.
The master equation technique includes the creation of atoms from the condensate and the annihilation of non-condensate particles through the emission of quasi-particles and vice versa for a finite number of atoms in the trap which gives rise to some thermal modification to the atom number fluctuations~\cite{Scully_PRA2000,Kocharovsky2006,Scully_2006,scully_2010,scully_PRL}.
To this end, we use the numerical results for the non-condensate particles $\tilde n$ and atom number fluctuations $\delta\tilde{n}^2$ extracted from Bogoliubov Hamiltonian (achieved from Eq.~\eqref{total_bogo.eqn}) and calculate the additional terms coming from the master equation technique which is proportional to the probability of finding different particle numbers in the condensate state. This modification which is an additional term to the non-condensate atom number fluctuation would directly modify the thermal behavior of the fluctuations of the atom number and helps to get a more accurate physical result.

In the following, we review the master equation technique proposed in Ref.~\cite{scully_2010} (known as CNB4) in order to clarify the atom number fluctuations result.
The equation of motion of the condensate particle in master equation formalism in terms of non-condensate particles and their fluctuations is given by

\begin{align}
    \frac{1}{k} \dot{P}_{N_0}=&-(N-N_0)(1+ \tilde{\mathbb{N}})(N_0+1)P_{N_0}
    \\& +(N-N_0+1)(1+\tilde{\mathbb{N}})N_0 P_{N_0-1} \nonumber
    \\& -N_0  (\mathcal{H}_I+(N-N_0)\tilde{\mathbb{N}}) P_{N_0} \nonumber 
    \\& +(N_0+1)(\mathcal{H}_I+(N-N_0-1)\tilde{\mathbb{N}}) P_{N_0+1}. \nonumber
\end{align}

Here $N$ is the total number of particles and $N_0$ is the number of particles in the condensate and the non-condensate particles or interaction term in terms of Bogoliubov quasi-particles in real space  is given by

\begin{align}
    \label{S:ntilde.eqn}
      \mathcal{H}_I &=\tilde{N}=\sum_i\Tilde{n}_{i,f} =\sum_{i,f}\langle\Tilde{\psi}_{i,f}^\dagger \Tilde{\psi}_{i,f} \rangle
      \\&
      =\sum_{b,i,f} (|u_{ib}|^2+|v_{ib}|^2 )N(E_{b} + f \hbar \omega_z)+|v_{ib}|^2.\nonumber
\end{align}

Here we define $\tilde{\mathbb{N}}$ in terms of the fluctuation of the non-condensate particles
\begin{equation}
    1+\tilde{\mathbb{N}}=\Delta \tilde{N}^2/\mathcal{H}_I=\Delta \tilde{N}^2/\tilde{N},
\end{equation}
with $\Delta \tilde{N}^2= \sum_{i,f} \delta\tilde{n}_{i,f}^2= \sum_{i,f} \langle\Tilde{\psi}^\dagger_{i,f} \Tilde{\psi}_{i,f} \Tilde{\psi}^\dagger_{i,f} \Tilde{\psi}_{i,f}\rangle-\langle\Tilde{\psi}^\dagger_{i,f} \Tilde{\psi}_{i,f}\rangle^2$.
The steady-state solution of the master equation above yields the probability distribution for the number of condensate particles, given by

\begin{align}
       P_{N_0}=\frac{(N-N_0-1+\mathcal{H}_I/\tilde{\mathbb{N}})!}{\mathcal{Z}_N (\mathcal{H}_I/\tilde{\mathbb{N}}-1)! (N-N_0)!} \left( \frac{\tilde{\mathbb{N}}}{\tilde{\mathbb{N}}+1}\right)^{N-N_0},
\end{align}
 with 
\begin{align}
       \mathcal{Z}_{N}=\sum_{N_0=0}^N
       \begin{pmatrix}
       N-N_0-1+\mathcal{H}_I/\tilde{\mathbb{N}} \\
       N-N_0
       \end{pmatrix}
        \left( \frac{\tilde{\mathbb{N}}}{\tilde{\mathbb{N}}+1}\right)^{N-N_0}.
\end{align}
Using these equations we can find condensate particles $N_0$ and their fluctuations $\Delta N_0^{2}$ using following relations.

\begin{align}
    & N_0=N-\mathcal{H}_I+P_{N_0}(\tilde{\mathbb{N}} N+\mathcal{H}_I)
    \\ &
    \Delta N_0^{2}=(1+\tilde{\mathbb{N}})\mathcal{H}_I-P_{N_0} (\tilde{\mathbb{N}} N +\mathcal{H}_I) (N+\tilde{\mathbb{N}}-\mathcal{H}_I+1)
    \nonumber \\ &
    -P_{N_0}^2 (\tilde{\mathbb{N}} N+\mathcal{H}_I)^2
\end{align}

\end{document}